\documentclass[preprint1]{aastex}

\shorttitle{Wang, et al.}
\shortauthors{SN 1987A}

\def\ApJ{ApJ}
\def\ApJS{ApJS}
\def\MNRAS{MNRAS}
\def\PASP{PASP}
\def\PASAU{PASAU}
\def\Nat{Nat}
\def\degree{$^{\rm o}$}
\def\kms{km s$^{-1}$}
\def\AsA{A\&A}

\begin{document}

\title{The Axially Symmetric Ejecta of Supernova 1987A}

\author{L.~Wang$^1$, J.~C.~Wheeler$^{2}$, P.~H\"oflich$^2$, A.~Khokhlov$^3$,
D.~Baade$^4$, D.~Branch$^5$, P.~Challis$^6$, A.~V.~Filippenko$^7$,
C.~Fransson$^8$, P.~Garnavich$^9$, R.~P.~Kirshner$^6$,
 P.~Lundqvist$^8$, R.~McCray$^{10}$, N.~Panagia$^{11}$, C.~S.~J. Pun$^{12}$,
M.~M.~Phillips$^{13}$, G.~Sonneborn$^{12}$,
N.B.~Suntzeff$^{14}$}

\affil{$^1$ Lawrence Berkeley Laboratory 50-232, 1 Cyclotron Rd,
Berkeley, CA~94720}
\affil{$^2$ Department of Astronomy, University of Texas, Austin, TX~78712}
\affil{$^3$ Laboratory for Computational Physics and Fluid Dynamics, Naval
Research Laboratory,
     Washington, DC~20375}
\affil{$^4$ European Southern Observatory, Karl-Schwarzschild-Strasse 2,
D-85748 Garching~bei~M\"unchen,
Germany}
\affil{$^5$ Department of Physics and Astronomy, University of Oklahoma,
Room 131 Nielsen Hall, Norman, OK~73019-0225.}
\affil{$^6$ Harvard-Smithsonian Center for Astrophysics, 60 Garden Street,
Cambridge, MA~02138 }
\affil{$^7$ Department of Astronomy, University of California, Berkeley,
CA~94720-3411}
\affil{$^8$ Stockholm Observatory, AlbaNova, Department of Astronomy,
SE-106 91 Stockholm, Sweden}
\affil{$^9$ Department of Physics, University of Notre Dame, Notre Dame,
IN~46556}
\affil{$^{10}$ Joint Institute for Laboratory Astrophysics, University of
Colorado, Boulder, CO~80309-0440}
\affil{$^{11}$ Space Telescope Science Institute, 3700 San Martin Drive,
Baltimore, MD~21218; on assignment from the Space Science
Department of ESA}
\affil{$^{12}$ NASA Goddard Space Flight Center, Laboratory for Astronomy
and Space Physics, Code 681, Greenbelt, MD~20771}
\affil{$^{13}$ Las Campanas Observatory, Carnegie Observatories, Casilla
601, La Serena, Chile}
\affil{$^{14}$ Cerro Tololo Inter-American Observatory, NOAO, Casilla 603,
La Serena, Chile}

\begin{abstract}
Extensive early observations proved that the ejecta of
supernova 1987A (SN 1987A) are aspherical. The
most important of these early observations include (a) the
``Bochum event" that revealed small-scale spectroscopic
structure indicating chemical inhomogeneities of the ejecta; (b)
spectropolarimetry that
showed deviations from spherical symmetry; (c) speckle
observations that revealed both the asymmetry of the ejecta
and the ``mystery  spot" manifested as
a secondary source off center from the bulk of the supernova
ejecta. Fifteen years after the  supernova explosion, the
\textit{Hubble Space Telescope} has resolved the rapidly expanding
ejecta.  The late-time images and spectroscopy provide
a geometrical picture that is consistent with early observations
and suggests a highly structured, axially symmetric geometry.
We present here a new synthesis of the old and new data. We show
that the Bochum event, presumably a clump of $^{56}$Ni, and
the late-time image, the locus of excitation by $^{44}$Ti,
are most naturally accounted for by sharing a common position
angle of about 14\degree, the same as the mystery spot and
early speckle data on the ejecta, and that they are both
oriented along the axis of the inner circumstellar ring at 45\degree\ to
the plane of the sky. We also demonstrate that the polarization
represents a prolate geometry with the same position angle and
axis as the early speckle data and the late-time image and hence
that the geometry has been fixed in time and throughout the ejecta.
The Bochum event and the Doppler kinematics
of the [Ca II]/[O II] emission in spatially resolved \textit{HST} 
spectra of the ejecta  can be consistently integrated into this geometry.
The radioactive clump is deduced to fall approximately along the axis of
the inner circumstellar ring and therefore to be redshifted in the North
whereas the [Ca II]/[O II] 7300 \AA\ emission is redshifted in the South.
We present a jet-induced model for the explosion and argue that such
a model can account for many of the observed asymmetries.
In the jet models, the oxygen and calcium are not expected to be 
distributed along the jet, but primarily in an expanding torus that 
shares the plane and northern blue shift of the inner circumstellar ring.
\end{abstract}

\section{Introduction}
  
SN 1987A is a Type II supernova in which the central iron core
of a massive star collapsed to form a dense neutron star or
black hole. Most of the conventional models assume that the
explosion process is roughly spherical.  The model assumption of
sphericity can be tested by observations.
It has been shown that core-collapse supernovae are
normally linearly polarized at a level around 1\% (Trammell
et al. 1993; Wang et al. 1996; Tran et al. 1997; Wang et al.
2001; Leonard et al. 2000, 2001; Leonard \& Filippenko 2001).
For a given supernova, the degree of polarization evolves with
time in a way that the  closer we see into the center the
higher the degree of  polarization (Wang et al. 1996; Wang et
al. 2001; Leonard et al. 2001). The degree of polarization
seems  also to be correlated with the mass of  the stellar
envelope; more massive envelopes show a lower degree of
polarization (Wang et al. 2001; Leonard \& Filippenko 2001). The
linear polarization measures the  asphericity of the supernova
photosphere, and the polarization data suggest models in which
the central engine that drives a core-collapse
supernova explosion is highly aspherical (H\"oflich, Wheeler, \& Wang 1999;
Khokhlov \& H\"oflich 2000;  Wheeler, et al. 2000;
H\"oflich, Khohklov \& Wang 2001).

As the brightest and the best observed supernova of modern astronomy, SN
1987A
provides a unique test case of the geometry of a
core-collapse supernova. Recent \textit{Hubble Space Telecsope} images
and spectroscopy are able to resolve the ejecta and it is now
possible to re-visit SN 1987A and see if the early and late
observations can be integrated into a single self-consistent
picture. In the following, we will discuss  the observed structures around
SN 1987A from the outermost circumstellar material inward to the central
rapidly expanding ejecta and provide a scenerio for many
of the features in terms of a jet-induced explosion.

\section{The Circumstellar Material of SN 1987A}

The circumstellar material (CSM) of SN 1987A
is highly structured (Wampler et al. 1990;
Crotts \& Heathcote 1991). The nebulosity consists
of a diffuse bow-shock shaped structure called ``Napolean's Hat''
and several nebular rings (Wampler et al. 1990; Wang \& Wampler 1992).
The nebular rings 
were shown beautifully by images taken by the \textit{HST} (Plait et al.
1995;
Burrows et al. 1995; Panagia et al. 1996).
The bright inner nebular rings were found to be
enriched in helium and nitrogen and are believed to be material
lost by the progenitor star before the explosion (Wang 1991;
Wang \& Mazzali 1992; Lundqvist \& Fransson 1996).
Napolean's Hat is also
likely to be associated with the mass loss of the progenitor
star (Wang \& Wampler 1992). The innermost bright (CSM) ring
observed by the \textit{HST} reveals a symmetry axis of the nebulosity
at a position angle (PA) of 89\degree\ that is roughly
consistent with the symmetry axis of the Napolean's Hat nebula
(Burrows et al. 1995). Napolean's Hat and the circumstellar
rings were formed long before the explosion of SN 1987A, but
they define a geometry that extends deep into the core of the
supernova.

\section{The Supernova Ejecta}

 Interior to the smallest circumstellar ring are
the  ejecta of the supernova, the material of which is
expanding at velocities from about
40,000 km s$^{-1}$ in the outermost hydrogen envelope (Pun et al. 1995)
down to about 1,300 km s$^{-1}$
in the carbon/oxygen-rich core. The ejected material
was too compact at early times
for normal direct imaging, but the asymmetry of the ejecta
was revealed by speckle observations (Papaliolios et al. 1989)
beginning in June 1987, after the second optical maximum.
Extensive linear
spectropolarimetry measurements were obtained during the first
year after the supernova  explosion (e.g. Bailey 1988; Mendez
et al. 1988; Schwartz 1987; Jeffery 1991). The linear
polarization is produced by electron scattering (Shapiro \&
Sutherland 1982; H\"oflich 1991) and probes the
geometry of the photosphere (see, however, Wang
\& Wheeler 1996 for an alternative model involving an
hypothesized circumstellar dust clump).  A general increase of
the polarization from about 0.1\% 2 days after  explosion to
1.1\% at $\sim$200 days was observed.
The dominant polarization position angle (PA) defined
by the loci of the data points on the Q-U diagram
(Wang et al. 2001) was
nearly constant at 110\degree\ from early to late observations (cf.
Q-U plots of Bailey et al. 1988: Cropper et al. 1988).
Since the degree of linear polarization increases
with the degree of  asymmetry of the ejecta, these polarimetry
observations show that the  scattering surface in the ejecta
becomes  increasingly asymmetric as we probe deeper into the
core. The fact that the polarization PA is constant
implies that the photosphere of the supernova has a single well-defined axis
of symmetry. 

More evidence of deviations from spherical symmetry is provided by the
so-called 
``Bochum event.'' During days 20-100, the H$\alpha$ line exhibited two
additional emission-like features at a velocity of about 
4500 \kms\ nearly symmetric on both the blue and red side of
the H$\alpha$ line (Hanuschik et al. 1988; Phillips \& Heathecote 1989).
The emission components are most likely to be caused by $^{56}$Ni
clumps ejected during the explosion (Lucy 1988; Utrobin,
Chugai, \& Andronova 1995).
The existence of high-velocity $^{56}$Ni was also suggested by
the earlier-than-expected detection of X-rays
(Dotani et al. 1987; Sunyaev et al. 1987)
and $\gamma$-rays (Matz et al. 1988).

The line of sight velocity of the Bochum clumps can be measured 
from the Doppler shift 
of the spectral profiles and compared to the photospheric velocity
using arguments based on geometric obscuration to deduce the transverse
velocity and space velocity (Utrobin et al. 1995).
By assuming photospheric velocities of 2800 and 2300 \kms\ on
days 29  and 41, respectively, Utrobin, et al. (1995) estimated
the transverse velocity of the $^{56}$Ni clump to be about 2400
\kms\ and the angle between the line of sight and the vector of
the absolute velocity to be 31\degree.  These results depend on
the location of the photosphere since the velocities must be
such that at a given epoch the photosphere does not obscure the
clump.  The observed blueshift of the minima of the P-Cygni
profiles of various lines show a large range. For
example, on day 27.76, the velocities  are -7603, -7262, -5968,
-3136, -2891, -5107, -4337, -3673, -2869 \kms\ for H$\alpha$,
H$\beta$, H$\gamma$,  Fe II $\lambda$5018, Fe II $\lambda$5169,
Na I $\lambda$5893, Ba II $\lambda$4554, Ba II $\lambda$6142,
and Sc II $\lambda$5527, respectively (Phillips et al. 1988).
The Bochum event appeared on day 20 when the photosphere was at
a velocity  around 3800 \kms\ as indicated by the measured
velocity of the  weak Fe II $\lambda$5018 line. Detailed NLTE
spectral analysis and the Doppler shifts of weak lines gave the
photospheric velocity to be around 4300 \kms\ on day 29, and
3200 \kms\ on day 41 (Lucy 1988; H\"oflich 1988). 

These higher photospheric velocities mean that at a given epoch the
photosphere was at a larger radius than assumed by Utrobin et
al. and hence that the clump must be at a larger angle to the
line of sight to avoid obscuration by the photosphere.  We find
that  the motion of the $^{56}$Ni clump is broadly consistent
with being tilted by roughly 45\degree\ from the line of
sight.  Chugai (1992) had earlier estimated the transverse
velocity of the $^{56}$Ni clump to be from 4000 \kms\ to 5000
\kms\ with a  representive mean of $\sim$ 4500 \kms. Such a
transverse velocity is in agreement with the conclusion that
the $^{56}$Ni clump is  directed at 45\degree\ from the line of
sight. The 45\degree\ orientation we deduce for the $^{56}$Ni
clump is  close to the inclination of the symmetry axis of the
circumstellar rings. Such a location of the $^{56}$Ni clump is
also consistent with the profile of the Ni II 6.6 $\mu$m
line observed at day 640 (Colgan et al. 1994).  While we cannot
determine the PA of these
$^{56}$Ni clumps on the plane of the sky from the Doppler shift data alone,
the strong implication is that they share a common axis with
the circumstellar rings.

Recent \textit{HST} images are able to spatially resolve the ejecta. Figure
1 shows
an image taken in the F439W filter on 2000 June 11
(see also Jansen \& Jakobsen 2001). The ejecta are obviously
elongated at a PA of about 14$\pm$5\degree.
At such a late stage, the images
in this filter 
show a mixture of emission from Fe I and
Fe II, with some minor contributions from hydrogen Balmer lines
(Wang et al. 1996; Chugai et al. 1997).  The images display the locus of the
late-time
excitation by $^{44}$Ti and hence show that the late-time deposition is
asymmetric and substantially bipolar.
The PA of the ejecta is remarkably consistent with the
alignment derived from early speckle observations of the ejecta
(20$\pm$5\degree; Papaliolios et al. 1989) and for the mystery spot
(14$\pm$3\degree\ to 194$\pm$3\degree; Nisenson et al. 1987; Meikle,
Matcher, \&
Morgan 1987; Nisenson \& Papaliolios 1999). The major axis of the ejecta
lies roughly in the direction of the minor axis of the inner nebular
ring and of Napoleon's Hat, but is noticeably offset by about 15\degree.

It is natural and expected that the $^{56}$Ni clumps of the Bochum
event and the distribution of $^{44}$Ti that powers the late-time
ejecta be associated.  This is another argument that the
$^{56}$Ni clumps share the PA of the images and the mystery spot, and in
turn that these features all share the same orientation with respect
to the line of sight, about 45\degree\ to the plane of the sky, away
from the observer in the North.

\section{The Structure of the Ejecta}

The schematic structure shown in Figure 2 may consistently account for most
of the observed features summarized above. Figure 2a shows the perspective
from Earth.  Figure 2b shows the structure projected onto the plane
formed by the line of sight and the
line from the origin to the North. The ring plane is inclined by about
44\degree\
to the line of sight. The $^{56}$Ni clump that can consistently
explain the redshifted component of the Bochum event is shown in Figure 2b
as well.  Note that although this $^{56}$Ni clump can be consistently
located on the symmetry axis of the ejecta by arguing that the $^{56}$Ni
and $^{44}$Ti have similar distributions, the only direct observational
constraint is that the $^{56}$Ni clump is located at an angle to the line
of sight of $\sim$ 45\degree.

To test whether or not the ejecta morphology observed with the \textit{HST}
is consistent with early spectropolarimetry,
we estimated the expected polarization
PA assuming either an oblate or prolate spheroid
for the scattering geometry.
We examined two cases, one with the
symmetry axis of the spheroid placed along the symmetry
axis of the inner ring and one with the symmetry axis
along the axis of the \textit{HST} image
of the ejecta (and of the mystery spot).
For the oblate case, the PA of the polarization is
co-aligned with the symmetry axis of the ellipsoid; for the prolate
case, the symmetry axis implied by the polarization is at 90\degree\
from the PA of the polarization.
Table 1 lists the PAs
of the major axis of the inner circumstellar ring and of the ejecta
and the predicted polarization PAs for
the two cases of oblate and prolate ellipsoids.
The polarimetry position angle defined by early broad-band observations
from Cropper et al. (1987) is 129\degree$\pm6.5$, and from Bailey (1988)
is 110\degree,
whereas the spectropolarimetry on June 3 for the 5800 \AA-region defines
a line at close to 0\degree\ (Cropper et al. 1988).  When all the
spectropolarimetric data are plotted on a single Q-U digram they define a
dominant axis corresponding to a position angle of
16.5\degree--106.5\degree$\pm3$ (Cropper et al. 1988). The
polarization PA corrected for interstellar polarization is around
110\degree; 
this is consistent only with a prolate spheroid oriented in the
same direction as the \textit{HST} images of the ejecta.

This is a remarkable concordance.  The early spectropolarimetry probes the
geometry of the outer hydrogen envelope, while the \textit{HST} images show
emission primarily from metal lines, a completely different region. The fact
that these different measures show similar structure is a
strong indication that both observations reveal the geometry of the ejecta,
despite
the fact that the ejecta may be heavily obscured by dust clumps synthesized
about
530 days after explosion (Lucy et al. 1989). In particular, the constancy of
the
polarization PA shows that the scattering geometry did not,
for instance, evolve from oblate to prolate.  The simplest hypothesis
is that the constant polarization angle reflected the geometry of
the ejecta ultimately revealed by the direct images of the ejecta.  This
hypothesis is confirmed by this quantitive check on the polarization
PA.  From the earliest epochs, the polarization of SN 1987A
revealed a prolate scattering geometry that was congruent with
the physical asymmetry of the iron-rich ejecta.

\section{Kinematics of the Ejecta}

At first glance, if the schematic picture of Figure 2b is true, we should
expect the northern part of the ejecta which points away from us to be
redshifted and the southern part of the ejecta to be blueshifted since it
points toward us. Spatially resolved spectroscopy obtained using the
\textit{HST} shows the opposite trend. Figure 3 shows the profile of
the [O II], [Ca II] 7300 \AA\ blend. The line peak of the southern part 
(the bottom slice) is obviously redshifted compared to that of the northern 
part (the top slice) by about 1270 \kms.

This apparent inconsistency has two important lessons to teach.  In the
first place, this pattern is not consistent with a spherical geometry
for the ejecta.  Even
with the effects of dust, one would expect the same Doppler shift in
the north and in the south, not a gradient, as seen.  We note that the
shift of line centers in Figure 3 is not monotonic from the north to
south, but the central slice, representing (within positioning errors,
see below) the centroid of the geometry, is blueshifted in an absolute
sense and with respect to the general trend from north to south. 
This departure from the trend could easily be explained by a clump
of obscuring dust that is thickest near the center of the ejecta.

The second lesson of Figure 3 is that the apparent discrepancy with the 
kinematics of the jet-like structure in Figure 2 can be resolved in 
the following way.  The ejecta are likely to be neither spherically 
symmetric nor homogeneously mixed and the [O II],
[Ca II] emission is probably formed in a different zone compared to 
the Fe I and Fe II emitting zones that are responsible
for the shape of the direct images. In particular, if the density of the
oxygen and calcium-rich ejecta is enhanced in the equatorial plane, then the
emissivity in the equatorial plane will be higher than at the poles.
This effect, plus some central dust, can easily produce the gradient 
in velocity from the north to south.  This behavior is expected
from a dusty oblate ellipsoid or torus.  Detailed analyses
of the spatially resolved spectra is clearly needed to map the ejecta
geometry and kinematics and the dust distribution, but is beyond the 
scope of this paper.  Nevertheless, we show in the next section that the 
inferred geometry -- nickel clumps along the axis and an expanding 
equatorial torus of oxygen and calcium --  are predicted by
a jet-induced model for the explosion.

While an equatorial torus is consistent with the overall gradient in
velocity displayed in Figure 3, we note the curious fact that the
whole system seems to have a rather large positive velocity (with
the exception of the possibly dust-obscured central portion).  The
southernmost portion has a velocity of about 1770 \kms\ and
the northernmost about 500 \kms.  A small portion of this overall
redshift can be accounted for by the space motion of the LMC of
290 \kms, but that leaves about 1000 \kms to explain.  Some of
this, but surely not the whole amount, could be due to a space motion
of the supernova itself.  We suspect that the bulk of the effect
is related to the fact that  
the feature in Figure 3 is actually a blend of [O II] $\lambda$7320 
and [Ca II] $\lambda$7300. It is not clear which is more dominant, 
so the rest frame wavelength (7300 \AA) used in making the plot may 
not be correct. A 20 \AA\ uncertainty corresponds to about
800 \kms.  There may also be some error in the STIS wavelength 
calibrations.  We have checked this hypothesis by examining the 
He I/Na I d blend.   If this feature is
dominated by Na D, then the data agree with a torus picture with
a blueshift in the north and a redshift in the south, but there is
a net offset if the feature is identified as pure helium.  
Another contribution could come from 
the uncertain positioning of the slit.  The overall
expansion speed of the inner portion of the ejecta (note the width
of the features in Figure 3 is about 3000 \kms.  While an attempt
was made to position the slit so that it was centered on the ejecta,
an offset by only one or two pixels to the south could mean that
the light down the slit was weighted toward southern, redshifted ejecta. 

The important aspect of these observations in the current context
is the clear gradient in velocity from north to south. In the
next section we show that this is consistent with the expectations
of a jet-induced explosion.

\section{Implications of Explosion Models of Core-Collapse Supernovae}

The observed bipolar structure of the ejecta and the ejection of
$^{56}$Ni at velocities above 5,000 km s$^{-1}$ argue
strongly that the engine that caused the supernova event was highly
aspherical. A model assuming  bipolar, non-relativistic jets formed at
the center of the explosion that can successfully eject the supernova
envelope can help us to understand how the asymmetry was produced
(Khokhlov et al. 1999; Khokhlov \& H\"oflich 2000; H\"oflich, Khokhlov, \&
Wang 2001).

 As an example, in Figure 4 we give the resulting chemical distribution of
the inner layers at 250 s after core collapse for an explosion of a
Type II supernova triggered by a non-relativistic jet with a velocity of
11,000 \kms\ that operates over $\sim 2.5$ and provides an explosion
energy of about $2 \times 10^{51}$ erg in a progenitor of 15 $M_\odot$.
At about 200 s, the jet material ``stalls'' in the expanding material. 
The density contours in the subsequent phases of the explosion
become almost spherical with the asymmetric chemical
profiles frozen into the envelope (Khokhlov \& H\"oflich 2000; H\"oflich,
Khokhlov, \& Wang 2001). During the phase of free expansion, 
the bow shock shows velocities of about 5000 \kms.  The resulting 
chemical composition structures are highly aspherical. 
Plumes of $^{56}$Ni will be selectively ejected along the jet axes 
and helium, oxygen, and calcium (not shown)
will be concentrated in expanding tori on the equator.

  The nickel plumes can possibly explain
the Bochum event and the apparent prolate asymmetry of the ejecta image.
In addition, the axial distribution of excitation
by radioactive decay will produce
an aspherical photosphere even in a nearly spherical density
distribution (Wang et al. 2001; H\"oflich, Khokhlov, \& Wang 2001).
Thus if $^{56}$Ni is distributed along a bipolar axis,
we expect the polarization to reflect that axis with an orientation
fixed in time, as observed.  The higher density of calcium in the
expanding equatorial torus will give a greater emission measure near
the equator and hence account for the observed northern blueshifted
component.

We feel it is highly
unlikely that a basically spherical model involving only
Rayleigh-Taylor instabilities could reproduce
the observed geometry. Jet-driven models provide an
interesting alternative, although the mechanism of jet formation during the
core-collapse process remains unclear.
Possible mechanisms of jet formation during the
collapse to form a neutron star have been discussed recently
(Wheeler et al. 2000, 2002; Akiyama et al. 2002) as have asymmetries
associated
with neutrino flow (Shimizu, Yamada, \& Sato 1994; Shimizu et al. 2001;
Fryer \& Heger 2000). Black-hole formation is another possibility
(MacFadyen
\& Woosley 1999). 

\section{Conclusions}

We have shown that SN 1987A has revealed a common bipolar axis of
symmetry from the early photospheric phase to the late imaging of the
ejecta.  We have shown for the first time that the images and the
polarimetry are consistent with a prolate geometry oriented nearly
(but not exactly) along the axis of the inner CSM ring.  We
have also deduced the orientation of the nickel clumps that were observed
as the ``Bochum event" with respect to the line of sight and argue that
these clumps are likely to be co-aligned with the bipolar ejecta
revealed in late \textit{HST} images.  The matter that shows up in the
images 
is precisely that heated by the long-term input of radioactive decay
of $^{44}$Ti.  Nagataki (2000) argues that the required abundance of
$^{44}$Ti itself may be evidence for explosion by a jet-like phenomenon.
Emission from [O II] and [Ca II] does not contribute substantially to the
\textit{HST} images,  but is shown in resolved spectroscopy to be
blueshifted in
the North  where the bipolar flow and one component of the Bochum event are
directed  away from the observer.  The existence and orientation of the
Bochum
event,  the common axis of polarization PA with the late-time image, and the
northern blueshifted oxygen and calcium are all consistent with, and
even predicted by, jet-induced supernova models.  In these models,
radioactive matter is ejected primarily along the poles and elements
synthesized 
in the progenitor (He, O, Ca) are ejected in expanding equatorial tori by
shocks
that propagate away from the bow shock of the jets and converge on the
equator.
The synthesis of observations we have presented here is consistent with a
model in which SN 1987A underwent a bipolar jet-induced explosion along a
common
axis that ran from the photosphere deep into the ejecta.

The small difference of 15\degree\ between the symmetry axis of the ejecta
and the inner CSM ring may provide some additional clues
to the overall evolution and explosion of SN~1987A. If the
axes of symmetry are all related to the rotation of the progenitor, then
we should expect a single axis during all these processes in order to
conserve angular momentum. What has changed the symmetry axis, and therefore
carried away angular momentum from the system?  Could it be that
the explosion was dominated by a magnetic field and
the inner ring was dominated by rotation?  Or vice versa?  Finally, we note
again
that the mystery spot closely aligns with the ejecta and the prolate axis of
the polarization geometry.  The mystery spot still begs explanation
(Rees 1987; Piran \& Nakamura 1987).  Perhaps the geometry we present
here can provide new clues to that problem as well.
These are all open questions to be pursued with
the strong suspicion that SN~1987A was a bipolar explosion.

\begin{center}
\textbf{Acknowledgements }
\end{center}
Support for this work was provided by NASA through grants GO 08243 and 09114
from the Space Telescope Science Institute, which is operated by the
Association 
of Universities for Research in Astronomy, Inc., under NASA contract
NAS5-26255.
Additional support was provided by NASA through grant NAG5-10766 and by
NSF through grant AST-0098644 to the University of Texas.

\newpage

\begin{table}[h]
Table 1. Polarization Position Angles \\

\begin{tabular}{l|l|l} \hline\hline

Geometry & P A of Major Symmetry Axis & Predicted P A of Pol. \\\hline

Oblate &179\degree (CSM ring) & 179\degree \\\hline
Prolate&179\degree (CSM ring) & 89\degree \\\hline

Oblate & 194\degree$\pm$5\degree\ (ejecta image)& 14$\pm$5\degree\ \\\hline
Prolate & 194\degree$\pm$5\degree\ (ejecta image)& 104$\pm$5\degree\ \\
\hline\hline

\end{tabular}

\end{table}

\newpage

\begin{figure}
\figurenum{1}
\epsscale{1.0}
\plotone{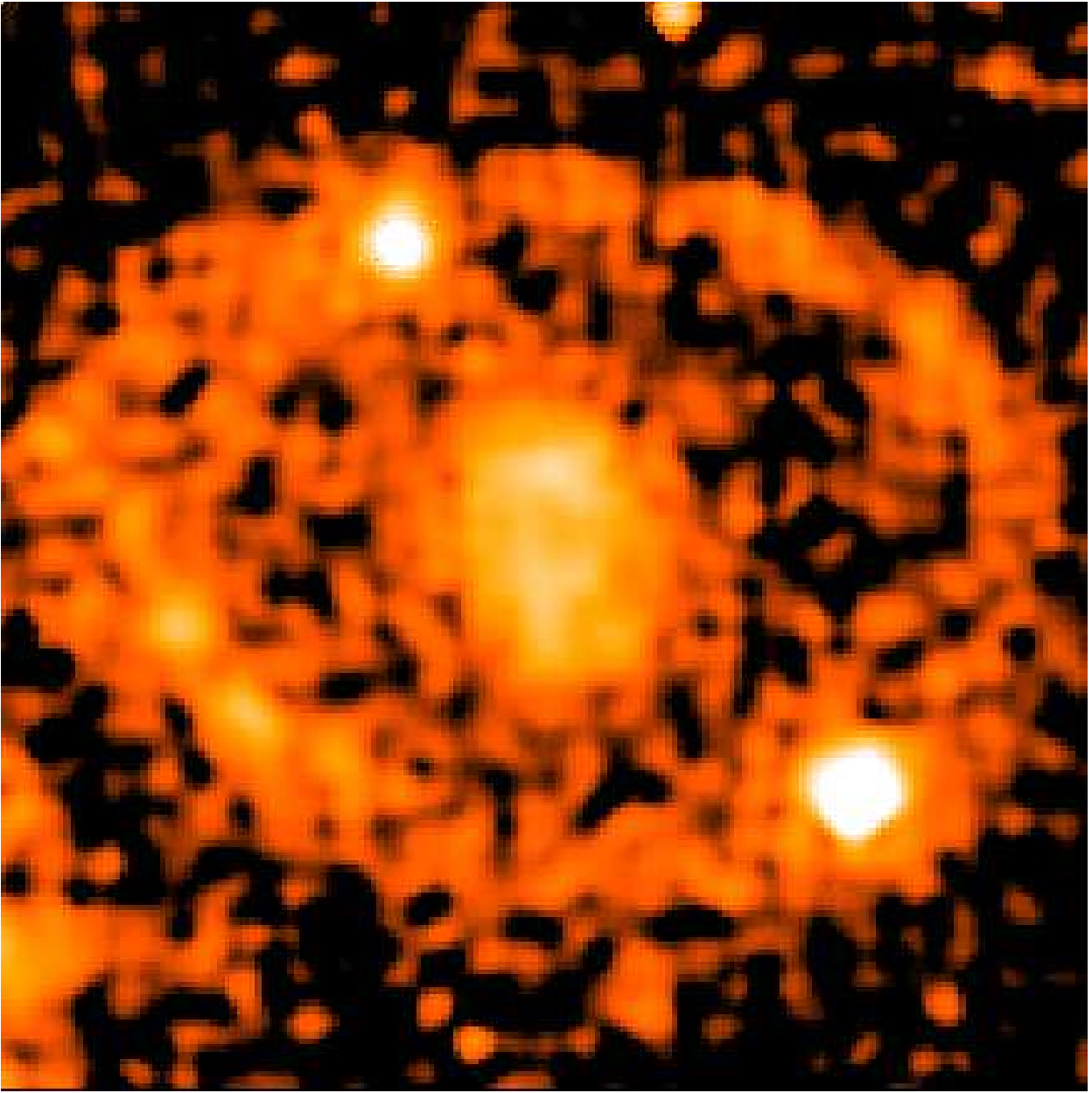}
\caption{\textit{HST} image taken on 2000 June 11 with filter F439W.
The apparent
asymmetry of the supernova ejecta represents the distribution of $^{56}$Fe.
East is to the left, North is up.
}
\end{figure}

\newpage

\begin{figure}
\figurenum{2}
\epsscale{1.0}
\plotone{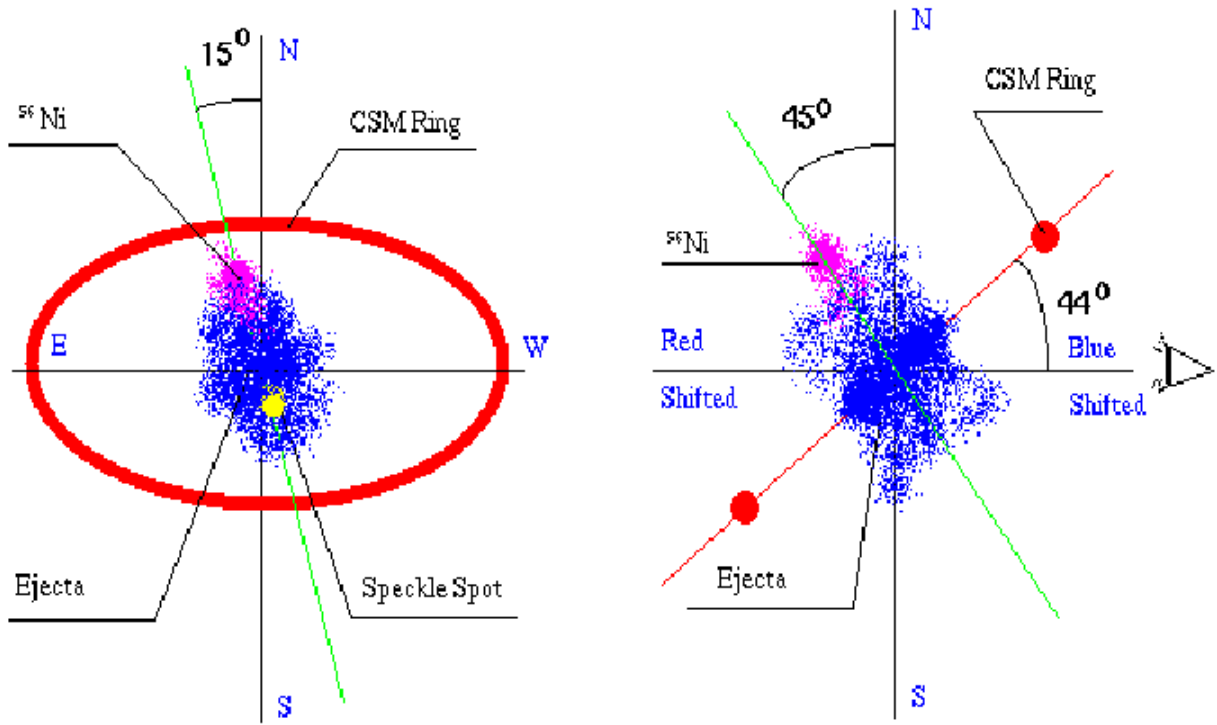}
\caption{Schematic picture of the inner CSM ring -- ejecta structure
projected onto (a) the plane of the sky, and (b) the plane of the line of
sight 
and the direction to the North.
The $^{56}$Ni clump at the northern tip of the
ejecta illustrates the so-called ``Bochum event." The  symmetry axis of the
ejecta is about 14\degree\ off to the east of the symmetry axis of the ring.
The $^{56}$Ni clump can be consistently placed at about 45\degree\ from the
line of sight.
}
\end{figure}

\newpage

\begin{figure}
\figurenum{3}
\epsscale{1.0}
\plotone{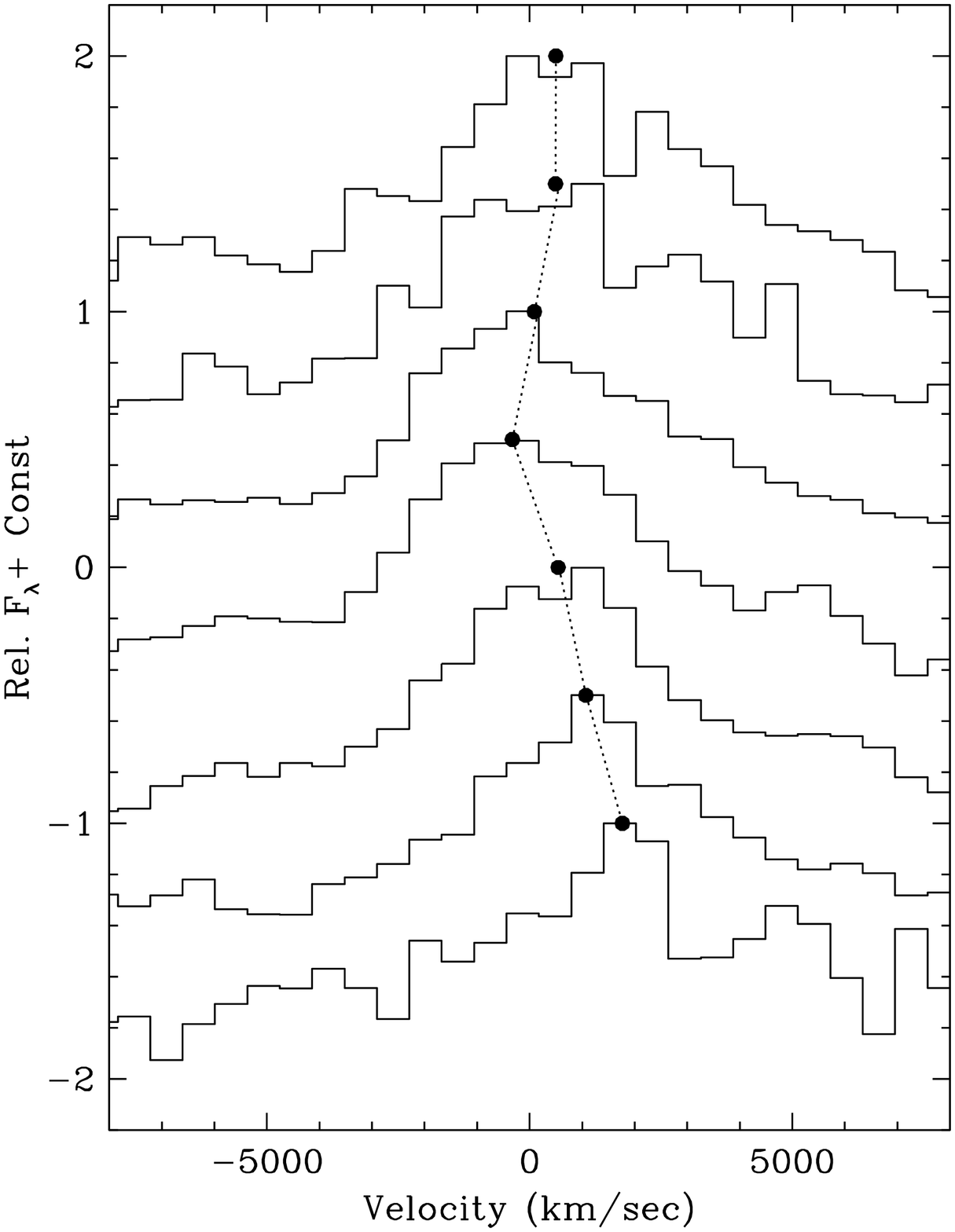}
\caption{HST-STIS velocity profile of the ejecta taken on
1999 Aug. 18 with the slit oriented nearly along the major axis of the ejecta.
From top to bottom, the solid lines show the [Ca II] 7300 \AA\ line in slices
from 0\arcsec.35 to 0\arcsec.25, from 0\arcsec.25 to 0\arcsec.15, 
from 0\arcsec.15 to 0\arcsec.05, from 0\arcsec.05 to -0\arcsec.05, 
from -0\arcsec.05 to -0\arcsec.15, from -0\arcsec.15 to -0\arcsec.25, and 
from -0\arcsec.25 to -0\arcsec.35 with positive numbers indicating to the 
north of the center of the ejecta. The solid circles denote the 
line centers. The spectra profiles are shifted 
vertically up arbitrarily for clarity. The peak of the northern side
of the ejecta is blueshifted compared to the southern side by about 1270 
\kms.  The behavior displayed here is not consistent with a spherically 
symmetric dusty sphere, for which the line peaks should 
have the same Doppler shift in the northernmost and the southernmost 
portions of the ejecta.  The overall gradient in velocity across the
ejecta from north to south is consistent with the schematic picture shown 
in Figure 2. The ejecta are obscured by (probably patchy) dust 
and thus line centers especially near the geometric center of the ejecta are 
shifted blueward compared to a dust-free configuration.  This may account
for the non-monotonic behavior of the line peaks near the center.  The
overall velocity of the ejecta may be affected by the placement of 
the slit and by the bulk motion of the supernova and the LMC (see text).
}
\end{figure}

\newpage
\begin{figure}
\figurenum{4}
\epsscale{1.0}
\plotone{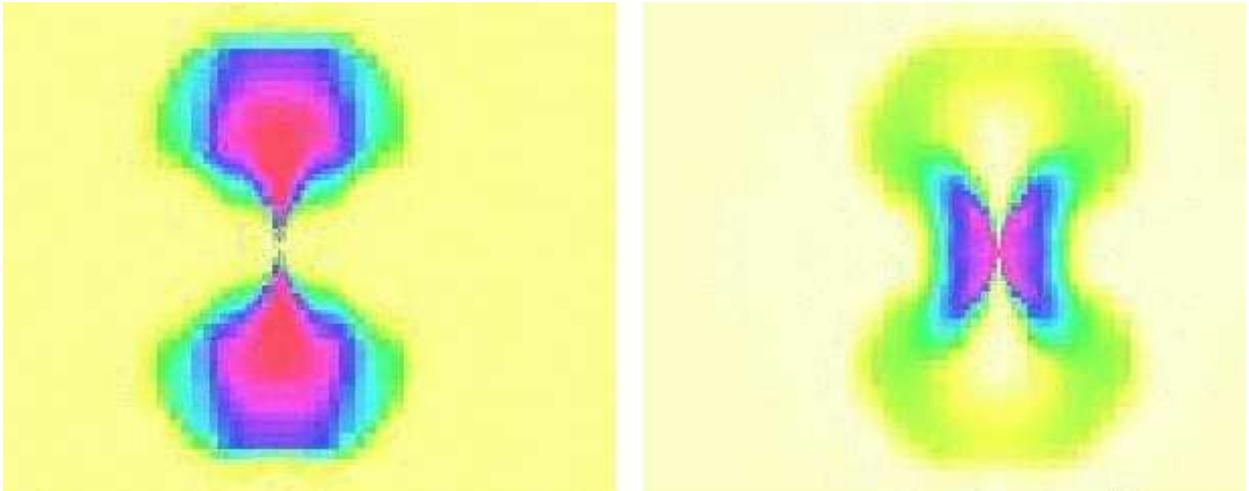}
\caption{Jet-driven hydrodynamic model of a core-collapse supernova.
 The distributions of elements  after ``freeze-out" are shown
for the jet-material (a; left) and for oxygen (b; right) in 
a ``snapshot" 250 s after explosion.  Each rectangle represents
a portion of the model $1.2\times10^12$ cm on a side at this epoch.  
The abundance mass fractions are color coded from yellow (0-25\%), green
(25-50\%), blue (50-75\%), to red (75-100\%).  Calcium is expected to have a
distribution similar to that of oxygen.  
}
\end{figure}

\end{document}